\documentstyle[12pt,epsf,epsfig]{article}
\def\ga{\mathrel{\raise.3ex\hbox{$>$\kern-.75em\lower1ex\hbox{$\sim$}}}}
\def\la{\mathrel{\raise.3ex\hbox{$<$\kern-.75em\lower1ex\hbox{$\sim$}}}}
\def\gyr{{\rm \, G\kern-0.125em yr}}
\def\gev{{\rm \, Ge\kern-0.125em V}}
\def\tev{{\rm \, Te\kern-0.125em V}}
\def\beq{\begin{equation}}
\def\eeq{\end{equation}}
\def\ss{\scriptscriptstyle}

\def\m12{m_{1\!/2}}

\def\mf{m_{\ss{f}}}

\def\thm{\theta_\mu}
\def\tha{\theta_A}

\def\cp{C\!P}

\begin{document}
\begin{titlepage}
\pagestyle{empty}
\baselineskip=21pt
\rightline{UMN--TH--1707/98}
\rightline{MADPH-98-1065}
\rightline{June 1998}
\vskip1.25in
\begin{center}

{\large{\bf New Contributions to Neutralino Elastic 
Cross-sections from \\
CP Violating Phases in the MSSM}}
\end{center}
\begin{center}
\vskip 0.5in
{Toby Falk}

{\it Department of Physics, University of Wisconsin, Madison, WI~53706,
WI, USA}

Andrew Ferstl and Keith A.~Olive

{\it
{School of Physics and Astronomy,
University of Minnesota, Minneapolis, MN 55455, USA}\\}
\vskip 0.5in
{\bf Abstract}
\end{center}
\baselineskip=18pt \noindent
We compute the four-Fermi neutralino-quark interaction Lagrangian
including contributions from the CP violating phases in the MSSM.
We find that neutralino-nucleus scattering cross-sections relevant for
direct detection experiments show a strong dependence on the value of
the CP-violating phase associated with the $\mu$ parameter, $\thm$.  In
some cases, for a broad range of non-zero $\thm$, there are cancellations
in the cross-sections which reduce the cross-section by more than an
order of magnitude.  In other cases, there may be enhancements as one
varies $\thm$.
\end{titlepage}
\baselineskip=18pt
The Minimal Supersymmetric Standard Model (MSSM) with a neutralino LSP
provides one of the better motivated candidates for the dark matter in the
Universe. From observations of the dynamics of galaxies and clusters of
galaxies \cite{gdyn}, and from the constraints on the baryon density from
big bang nucleosynthesis \cite{bbn}, it is clear that a considerable
amount of non-baryonic dark matter is needed. The MSSM, with supersymmetry
breaking mediated by gravitational interactions  and with R-Parity
conservation, typically possesses a stable dark matter candidate, the
LSP,  which for much of the parameter space is a neutralino (a linear
combination of the SU(2) and U(1) gauginos, and the two Higgsinos) with a
mass in the range $m_\chi \sim O(1-100) \gev$.
In fact, there has been considerable progress recently in establishing
strong constraints on the supersymmetric parameter space from recent runs
at LEP \cite{efos2,efgos}. These constraints provide a lower bound to the
neutralino mass of $\sim 40\gev$, when in addition to the bounds from
experimental searches for charginos, associated neutralino production and
Higgs bosons,  constraints coming from cosmology and theoretical
simplifications concerning the input scalar masses in the theory are
invoked. (The pure experimental bound is about $m_\chi \ga 30$ GeV
\cite{someexperiment}). 

A major issue concerning dark matter of any kind is
it detection and identification.  Indeed, there are a multitude of
ongoing experiments involved in the direct and indirect detection of
dark matter, many with a specific emphasis on searching for
supersymmetric dark matter \cite{jkg}.  The event rates for  either
direct or indirect detection depend crucially on the dark matter elastic
cross-section, in this case the neutralino-nucleon, or neutralino-nucleus,
cross-section.  Because the neutralinos have Majorana mass terms, their
interactions with matter are generally spin dependent, coming from an
effective interaction term of the form ${\bar \chi} \gamma^\mu \gamma^5
\chi {\bar q} \gamma_\mu \gamma^5 q$.   In the regions of the MSSM parameter space where the LSP is
a mixture of both gaugino and Higgsino components, or where the squarks
are highly mixed \cite{sw},
there is also an important contribution to the scattering
cross-section due to a term in the interaction Lagrangian of the form 
${\bar \chi} \chi {\bar q} q$ \cite{g} which is spin independent.  These
terms are particularly important for scattering off of large nuclei,
where coherent nucleon scattering effects can quickly come to dominate
all others.

When gaugino mass unification at the GUT scale is assumed, as is done
here, the identity of the LSP in the MSSM is determined by three parameters.
These are the gaugino mass, represented here as the SU(2) gaugino
mass $M_2$ at the weak scale; the Higgsino mixing mass, $\mu$; and the
ratio of Higgs vevs,
$\tan \beta$. The interactions of the LSP with matter also depend on
additional mass parameters, specifically the sfermion and Higgs masses, which in
turn are determined from the soft supersymmetry breaking sfermion masses,  trilinear
and bilinear parameters, $m_i, A_i$, and $B$.  It is very common to choose a common
soft sfermion mass $m_0$ at the GUT scale, which greatly reduces the
number of available parameters.  In some cases, the Higgs soft masses are
also chosen equal to the common sfermion soft masses at the GUT scale.
This assumption leads to what is known as the Constrained MSSM.  In the
CMSSM, two parameters, usually $\mu$ and the Higgs pseudo-scalar mass, are
fixed by the condition of proper electroweak symmetry breaking.  The CMSSM
generally leads to a nearly pure bino as the LSP, and as we want to
consider all neutralino compositions,  we will not
consider the case of universal soft Higgs masses, though for simplicity we will assume
that the remaining (sfermion) soft masses are unified at the GUT scale.

The MSSM is well known to contain several independent
$\cp$-violating phases. If one assumes that all of the supersymmetry breaking
trilinear mass terms, $A_i$, are equal to $A_0$ at the GUT scale, then
the number of independent phases reduces to 2, which one can take as
$\tha$ and
$\thm$. The phase of $\mu$ can always be adjusted so that it is
equal and opposite to that of the supersymmetry breaking bilinear mass
term $B$, $\theta_B = -
\thm$ by rotating the Higgs fields so that their vacuum expectation
values are real\footnote{Note that in some cases loop effects
may not allow this simple tree level rotation \cite{pil1}.}
\cite{dgh}.  Though these phases can lead to sizable contributions to the
neutron and electron dipole moments \cite{dn}, it has been shown that
large phases are indeed compatible with these constraints, as well as
cosmological constraints on the neutralino relic density in the MSSM
\cite{fkosi} and in the Constrained MSSM \cite{fko1,fko2}.   Indeed, in
the CMSSM,  cancellations between different contributions to the EDMs over
a broad range in mass parameters allow for  a $\thm$  as large as $\sim 0.3 \pi$,
depending on the magnitude of $A_0$ and $\tan \beta$, and a 
$\tha$ which is essentially unconstrained.
If we drop the assumption of universal Higgs masses at the GUT scale, these phases
are even less constrained.  

Here we will show the importance of the CP-violating phases on the
elastic scattering cross-sections of neutralinos on matter. To this
effect, we will calculate the four-Fermi $\chi$-quark interaction
Lagrangian with the inclusion of the CP violating phase $\thm$ for the
standard spin dependent and spin independent interactions.  Here, 
we have chosen ${\tha}_i = \pi/2 $ and adjusted the 
magnitude of the $A_i$ (of order 1 -- 3 TeV) in order to
satisfy the bounds for the electric dipole moments of the electron and
neutron. That this can be done has been demonstrated in \cite{gw} where
it was shown that the neutron electric dipole moment contains separate
contributions from the imaginary parts of $A_u$, $A_d$,
and $B$. Though this can be regarded as a fine-tuning, our purpose
here is to concentrate on the behavior of the elastic scattering
cross-section rather than the cancellation of the electric dipole
moments which has been treated at legnth elsewhere. 
A complete treatment
of the effective Lagrangian which includes $\tha$ as well as new
annihilation contributions for non-zero phases will be presented elsewhere
\cite{ffo2}.

Before writing down the effective Lagrangian, it will be useful to clarify our
notation.  We will write the lowest mass neutralino eigenstate (the LSP) as
\beq
\chi = {Z_\chi}_1 {\tilde B} + {Z_\chi}_2 {\tilde W} + {Z_\chi}_3 {\tilde
H}_1 + {Z_\chi}_4 {\tilde H}_2
\eeq
The neutralino mass matrix in the $({\tilde B}, {\tilde W}^3, {{\tilde
H}^0}_1,{{\tilde H}^0}_2 )$ basis
\beq
  \left( \begin{array}{cccc}
M_1 & 0 & {-M_Z \sin \theta_W \cos \beta} &  {M_Z \sin \theta_W \sin
\beta} \\ 0 & M_2 & {M_Z \cos \theta_W \cos \beta} & {-M_Z \cos \theta_W
\sin \beta}
\\ {-M_Z \sin \theta_W \cos \beta} & {M_Z \cos \theta_W \cos \beta} & 0
& -\mu
\\ {M_Z \sin \theta_W \sin \beta} & {-M_Z \cos \theta_W \sin \beta} &
-\mu & 0 
\end{array} \right) 
\label{mm}
\eeq
depends explicitly on the Higgsino mass
parameter $\mu$, and the coefficients ${Z_\chi}_i$ all depend on the phase
$\thm$.  In (\ref{mm}), we have taken $M_1 = {5 \over 3} \tan^2 \theta_W
M_2$.
 The phases could in principle
also enter into the calculation through the sfermion mass eigenstates.
The sfermion mass$^2$ matrix can be written as
\begin{equation}
\pmatrix{ M_L^2 + m_f^2 + \cos 2\beta (T_{3f} - Q_f\sin ^2 \theta_W) M_Z^2 &
-m_f\,\overline{m}_{\ss f} e^{i \gamma_f}
\cr
\noalign{\medskip} -m_f\,\overline{m}_{\ss f} e^{-i \gamma_f} & M_R^2 + m_f^2 +
\cos 2\beta Q_f\sin ^2
\theta_W M_Z^2
\cr }~
\label{sfm}
\end{equation}
where $M_{L(R)}$ are the soft supersymmetry breaking sfermion masses,
which we have assumed are generation independent and generation
diagonal and hence real.  Due to our choice of phases, there is a
non-trivial phase associated with the off-diagonal entries, which we
denote by $-\mf(\overline{m}_{\ss f} e^{i \gamma_f})$, of the sfermion
mass$^2$ matrix, and
\begin{equation}
  \label{mfbar}
  \overline{m}_{\ss f} e^{i \gamma_f} = R_f \mu + A_f^* = R_f
  |\mu|e^{i \theta_\mu} + |A_f|e^{-i \theta_{A_{\ss f}}},
\end{equation}
where $m_f$ is the mass of the fermion $f$ and $R_f =
\cot\beta\:(\tan\beta)$ for weak isospin +1/2 (-1/2) fermions.  We
also define the sfermion mixing angle $\theta_f$ by the unitary matrix
$U$ which diagonalizes the sfermion mass$^2$ matrix,
\begin{equation}
U = \pmatrix{ \cos\theta_{\!f} & \sin\theta_{\!f}\, e^{i\gamma_f} \cr
  \noalign{\medskip}
              -\sin\theta_{\!f} \,e^{-i\gamma_f} & \cos\theta_{\!f}\cr }
\equiv \pmatrix{ \eta_{11} & \eta_{12} \cr
  \noalign{\medskip}
              \eta_{21} & \eta_{22}\cr }.
\end{equation}
Note that $\eta_{21} = - \eta_{12}^*$.

The general form for the four-Fermi
effective Lagrangian can be written as
\beq
{\cal L} = {\bar \chi} \gamma^\mu \gamma^5 \chi {\bar q}_i \gamma_\mu
(\alpha_1 + \alpha_2 \gamma^5) q_i + \alpha_3  {\bar \chi} \chi {\bar
q}_i q_i + \alpha_4 {\bar \chi} \gamma^5\chi {\bar q}_i \gamma^5 q_i +
\alpha_5 {\bar
\chi} \chi {\bar q}_i \gamma^6 q_i + \alpha_6 {\bar \chi}  \gamma^5 
\chi {\bar q}_i q_i
\eeq
The Lagrangian should be  summed over quark generations, and the
subscript $i$ refers to up-type ${i=1}$ and down-type ${i=2}$ quarks.
Here, we shall only be concerned with the axial vector ($\alpha_2$) and
scalar ($\alpha_3$) contributions.  These coefficients
are given by:
\newpage
\begin{eqnarray}
{\alpha_2}_i & = &   {1 \over 4({m_1^2}_i - m_\chi^2)} \left[ \left|\eta_{11}^*
 \left({Y_i \over 2} g' {Z_\chi}_1 + g {T_3}_i {Z_\chi}_2\right) + {\eta_{12}^*
 g {m_q}_i {Z_\chi}_{5-i}  \over 2 m_W B_i}\right|^2  \right. \nonumber \\
& & \left. + \left|-\eta_{12}^* e_i
 g' {Z_\chi^*}_1 +  {\eta_{11}^* g {m_q}_i
{Z_\chi^*}_{5-i}  \over 2 m_W B_i}\right|^2 \right] \nonumber \\
& &  + {1 \over 4({m_2^2}_i - m_\chi^2)} \left[ \left|\eta_{21}^* \left({Y_i
\over 2} g' {Z_\chi}_1 + g {T_3}_i {Z_\chi}_2\right) + {\eta_{22}^* g
{m_q}_i {Z_\chi}_{5-i}  \over 2 m_W B_i}\right|^2 \right.\nonumber  \\
& & \left. + \left|-\eta_{22}^* e_i
 g' {Z_\chi^*}_1 +  {\eta_{21}^* g {m_q}_i
{Z_\chi^*}_{5-i}  \over 2 m_W B_i}\right|^2 \right] \nonumber \\
& & - {g^2 \over 8 m_Z^2 \cos^2 \theta_W}  (|{Z_\chi}_3|^2 -
|{Z_\chi}_4|^2) {T_3}_i
\label{a2}
\end{eqnarray}
\begin{eqnarray}
{\alpha_3}_i & = & - {1 \over 2 ({m_1^2}_i - m_\chi^2)} {\rm Re} \left[ 
\left( { \eta_{11}^* g {m_q}_i {Z_\chi}^*_{5-i}  \over 2 m_W B_i} - \eta_{12}^*
e_i g' {Z_\chi^*}_1 \right) \right.\nonumber  \\
 & & \left. \times \left( \eta_{11}^*\left({Y_i \over 2} g' {Z_\chi}_1 + g
{T_3}_i {Z_\chi}_2\right) + {\eta_{12}^* g {m_q}_i {Z_\chi}_{5-i} \over 2 m_W
B_i} \right)^* \right] \nonumber \\
& & - {1 \over 2 ({m_2^2}_i - m_\chi^2)} {\rm Re} \left[  \left(
{ \eta_{21}^* g {m_q}_i {Z_\chi}^*_{5-i}  \over 2 m_W B_i} - \eta_{22}^* e_i
 g' {Z_\chi^*}_1 \right) \right.\nonumber  \\
 & & \left. \times \left( \eta_{21}^*\left({Y_i \over 2} g' {Z_\chi}_1 + g
{T_3}_i {Z_\chi}_2\right) + {\eta_{22}^* g {m_q}_i {Z_\chi}_{5-i} \over 2 m_W
B_i} \right)^* \right] \nonumber \\
& & - {g m_{q_i} \over 4 m_W B_i} \left[ {\rm Re} \left( \delta_{1i} [
g  Z_{\chi_2} - g'  Z_{\chi_1} ] \right) C_i D_i \left( -{1 \over
m_{H_1}^2} + {1 \over m_{H_2}^2}\right) \right. \nonumber \\
& & \left. + {\rm Re} \left( \delta_{2i} [ g  Z_{\chi_2}
- g'  Z_{\chi_1} ] \right) \left( {D_i^2 \over m_{H_2}^2} + {C_i^2 \over
m_{H_1}^2}\right) \right]
\label{a3}
\end{eqnarray}
In these expressions, ${m_{1,2}}_i$ are the squark mass eigenvalues, $B_i =
\sin \beta (\cos \beta)$ for up (down) type quarks and
$C_i = \sin \alpha (\cos \alpha)$, 
$D_i = \cos \alpha (-\sin \alpha)$ ($\alpha$ is the scalar Higgs mixing
angle),  $\delta_{1i}$ is ${Z_\chi}_3$
(${Z_\chi}_4$), and $\delta_{2i}$ is ${Z_\chi}_4$ ($-{Z_\chi}_3$). 
In the limit of vanishing CP-violating phases, these expressions agree
with those in \cite{jkg} and \cite{ef}.   Expressions for
${\alpha_1}_i $, ${\alpha_4}_i $, ${\alpha_5}_i$ and 
${\alpha_6}_i $, which are suppressed by the neutralino-quark relative velocity, 
 will be presented in \cite{ffo2}.

Equations (\ref{a2}) and (\ref{a3}) contain
contributions to the effective Lagrangian for neutralino-quark scattering from squark, $Z$, and both
scalar Higgs exchange.  The spin dependent contribution (from $\alpha_2$)
contains terms which are not suppressed by the quark mass and can be large
over much of the parameter space, that is, they do not rely on the LSP being
a mixed gaugino-Higgsino
eigenstate, i.e. having both a large $Z_{\chi_{1,2}}$ and a large
$Z_{\chi_{3,4}}$ component. In contrast, the spin independent term (from
$\alpha_3$) is always proportional to the quark mass and relies on either
the LSP being a mixed state or significant squark mixing \cite{sw}.
However, the spin independent cross-section is enhanced by the effects
of coherent scattering in a nucleus and can dominate over the spin
dependent cross-section for heavy nuclei.

The elastic scattering cross sections based on $\alpha_{2,3}$ have been
conveniently expressed in \cite{jkg}.  The spin dependent cross-section
can be written as
\beq
\sigma_{2} = {32 \over \pi} G_F^2 m_r^2 \Lambda^2 J (J+1)
\eeq
where $m_r$ is the reduced neutralino-nucleus mass, $J$ is the spin of the
nucleus and 
\beq
\Lambda = { 1\over J} (a_p \langle S_p \rangle + a_n \langle S_n \rangle)
\eeq
and 
\beq
a_p = \sum_i {{\alpha_2}_i \over \sqrt{2} G_F} \Delta^{(p)}_i, \qquad 
a_n = \sum_i {{\alpha_2}_i \over \sqrt{2} G_F} \Delta^{(n)}_i
\label{as}
\eeq
The factors $\Delta^{(p,n)}_i$ depend on the spin content of the nucleon and
are taken here to be $\Delta^{(p)}_i = 0.77, -0.38, -0.09$ for $u,d,s$
respectively \cite{smc} and $\Delta^{(n)}_u = \Delta^{(p)}_d,
\Delta^{(n)}_d =  \Delta^{(p)}_u, \Delta^{(n)}_s = \Delta^{(p)}_s$. The
$\langle S_{p,n} \rangle$ are expectation values of the spin content in
the nucleus and therefore are quite dependent on the target nucleus.  We
will display results for scattering off of a $^{73}$Ge target for which in
the shell model $\langle S_{p,n}
\rangle = 0.011, 0.491$, and for $^{19}$F, which has  $\langle S_{p,n}
\rangle = 0.415, -0.047$. For details on the these
quantities, we refer the reader to \cite{jkg}. 

Similarly, we can write the spin independent cross section as
\beq
\sigma_{3}  = {4 m_r^2 \over \pi} \left[ Z f_p + (A-Z) f_n \right]^2
\eeq
where
\beq
{f_p \over m_p} = \sum_{q=u,d,s} f_{Tq}^{(p)} {\alpha_3}_q /m_q + 
{2 \over 27} f_{TG}^{(p)} \sum_{q=c,b,t} {\alpha_3}_q/m_q
\label{fp}
\eeq
and a similar expression for $f_n$. The parameters $f_{Tq}^{(p)}$ are
defined by $\langle p | m_q {\bar q} q | p \rangle = m_p f_{Tq}^{(p)}$,
while
$f_{TG}=1-(f_{Tu}+f_{Td}+f_{Ts})$ \cite{svz}.  We have
adopted $f_{Tq}^{(p)} = 0.019, 0.041, 0.14$ for $u,d,s$ and for 
$f_{Tq}^{(n)} = 0.023, 0.034, 0.14$ \cite{gls}.   
The cross-sections derived from (\ref{a3}) and 
(\ref{fp}) approximate the squark exchange contributions for heavy
quarks \cite{dn1} and neglect the effect of twist-2 operators;  however
the change from  a more careful treatment of loop effects for heavy
quarks and the inclusion of twist-2 operators is numerically small
\cite{dn2}.

We are now ready to show the importance of the phases. As we noted
earlier, we will restrict our parameter choices to universal gaugino
masses and universal sfermion masses at the GUT scale. We will also
choose $\tan\beta = 3$ throughout to make it easier to remain consistent
with recent constraints on the Higgs mass of about 78 GeV (for this value
of $\tan
\beta$) \cite{higgs}. We will also choose $m_0 = 100 \gev$ throughout.
Because of the running of the RGE, this leads to typical squark masses
of $\sim 450$ GeV for $M_2\sim 150$ GeV. Finally, we have chosen the
value of the pseudo-scalar Higgs mass to be 300 GeV. 

\begin{figure}
\begin{center}
\vspace*{-0.5in}
\hspace*{-0.0in}
\epsfig{file=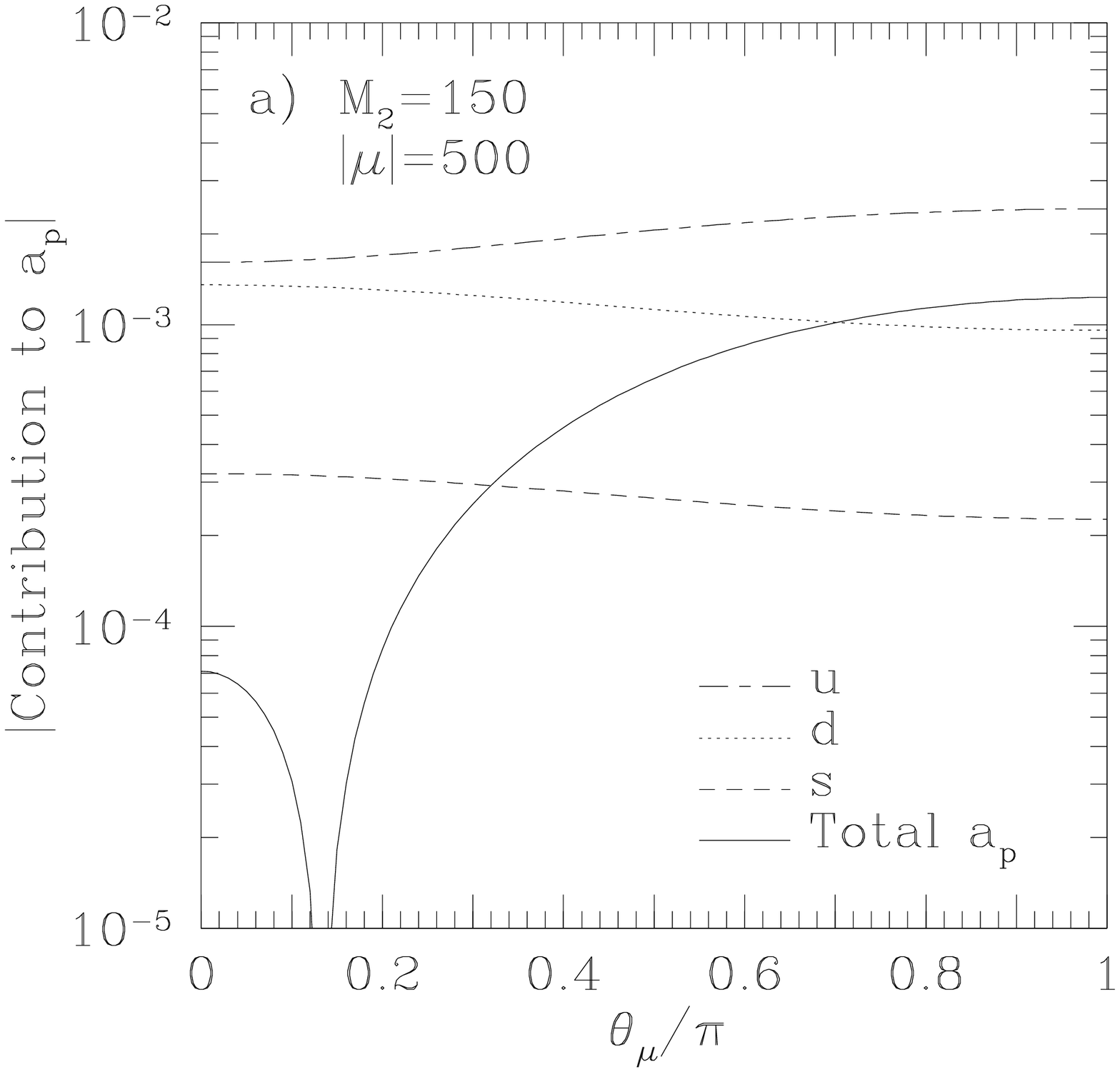,height=4.in} 
\epsfig{file=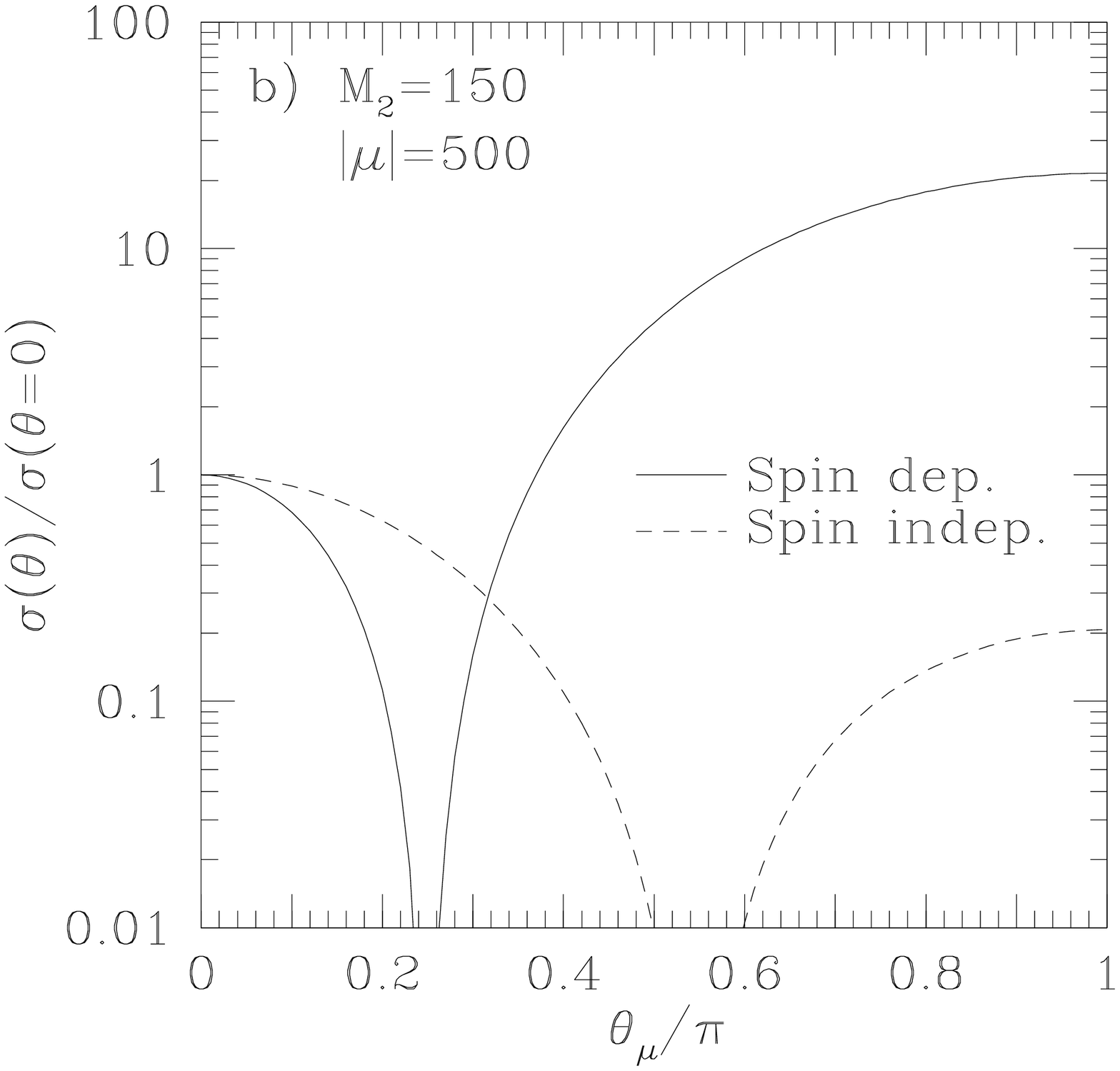,height=4.in} 
\caption{\label{fig:fl}For elastic scattering off of ${}^{19}$F,
a) the absolute value of the contributions to $a_p$ from
individual quarks as a function of
$\thm$, and b) spin dependent (solid) and spin independent (dashed)
cross-sections as a function
of $\thm$, and normalized to the cross-sections at $\thm=0$.}
\end{center}
\end{figure}

 We begin our
discussion by focusing on the spin dependent contribution from $\alpha_2$.
For this case, we consider the scattering of neutralinos on fluorine, for
which the spin dependent contribution typically dominates by a factor of
about 20 \cite{jkg}. In Fig.~\ref{fig:fl}a, we show the contributions from different quarks to
$a_p$ given in eq. (\ref{as}), as a function of the CP-violating phase
$\thm$, for $M_2=150$ GeV and  $\mu=500$ GeV.  In this case, both the
contributions from squark exchange and $Z$ exchange are significant.  The signs of the
individual $\alpha_{2i}$'s are all positive;  however, the sign of the contribution to 
$a_p$ is different for the $u$ quark than for the $d$ and $s$ quarks due to 
the different sign in  $\Delta_u$  relative to $\Delta_d$ and $\Delta_s$.
As one can see, there are important cancellations which can dramatically 
reduce the spin dependent 
cross-section. 
We note that since this sign difference in the $\Delta$'s is generic, this effect does
not depend heavily on the spin structure of the nucleon. 

The total value
of the spin dependent cross-section  $\sigma_2(\thm)$ for
$\chi$-$^{19}$F scattering is shown by the solid curve in Fig.~\ref{fig:fl}b,
normalized to the value of the spin dependent cross-section at $\thm = 0$. 
For these values of the MSSM parameters, the neutralino is predominantly a
${\tilde B}$ with a mass $m_\chi \simeq 75 \gev$. The relic density is
about $\Omega h^2 \simeq 0.15$. As one can
see, there is an important dependence on $\thm$ and a cancellation in 
$\sigma_2$ leading to a decrease in the cross-section by at least an
order of magnitude for
$\thm/\pi$ = 0.2 -- 0.3.  The presence of such a large cancellation
in the spin dependent cross-section 
over a range in $\thm$ and its position in $\thm$ depend on the MSSM parameters.
For comparison, we also show by the dashed
curve, the spin independent cross-section $\sigma_3$ for the same MSSM
parameters and for scattering on fluorine, which also exhibits a similar reduction
near $\thm/\pi$ = 0.5 -- 0.6.
 Note that the neutralino relic density is not strongly dependent on
$\thm$ since the ${\tilde B}$  mass is insensitive to $\mu$. Furthermore,
the ${\tilde B}$ relic density depends primarily on the annihilation
through slepton exchange (since squarks are heavier when universal
sfermion masses are assumed at the GUT scale and $m_0 \la M_2$). 
Because slepton mixing is small (the off-diagonal elements in (\ref{sfm})
are proportional to the lepton masses) the dependence on the CP-violating
phases is also small \cite{fko1}.

\begin{figure}
\begin{center}
\vspace*{-0.5in}
\hspace*{-0.0in}
\epsfig{file=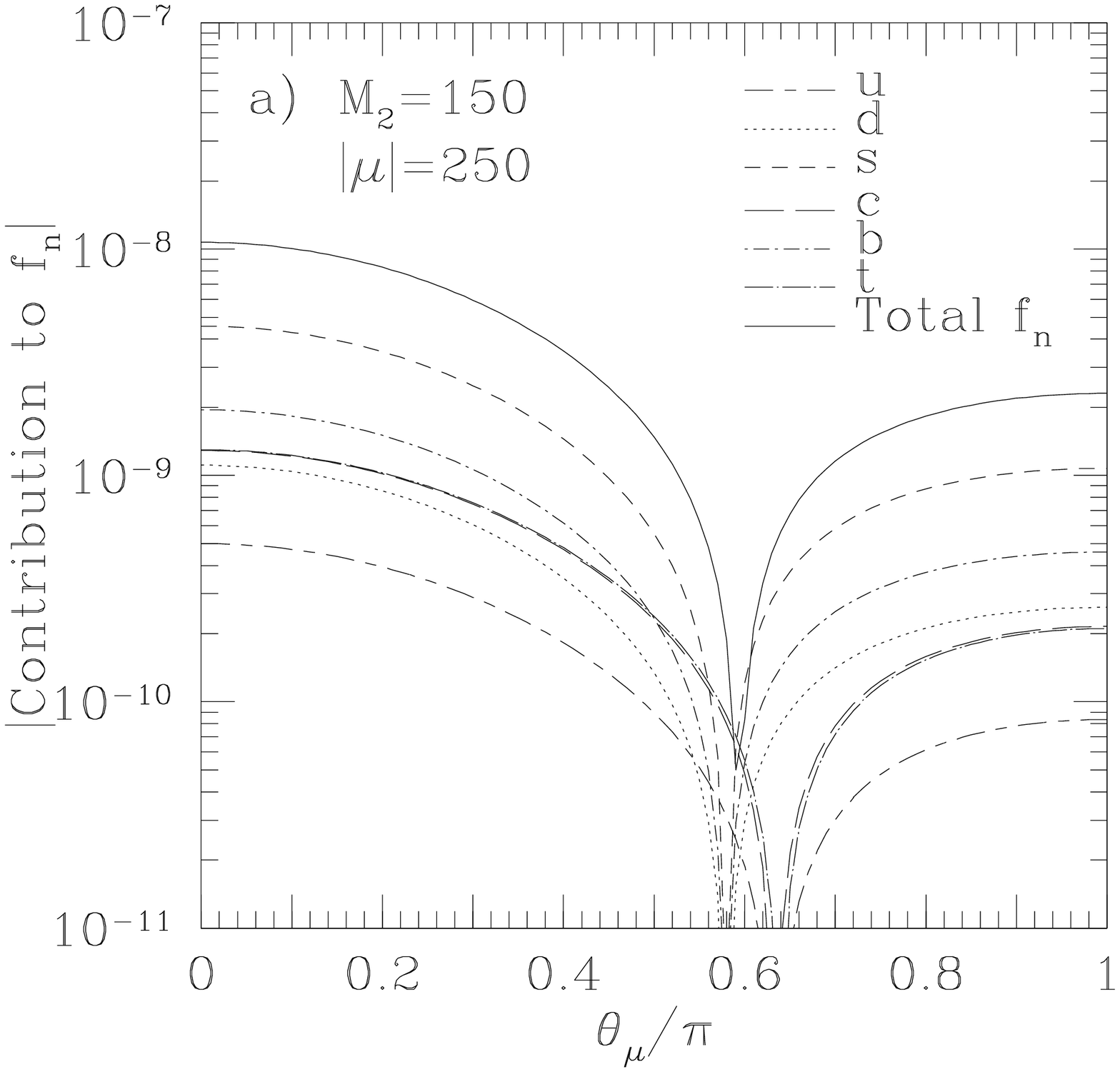,height=4.in} 
\epsfig{file=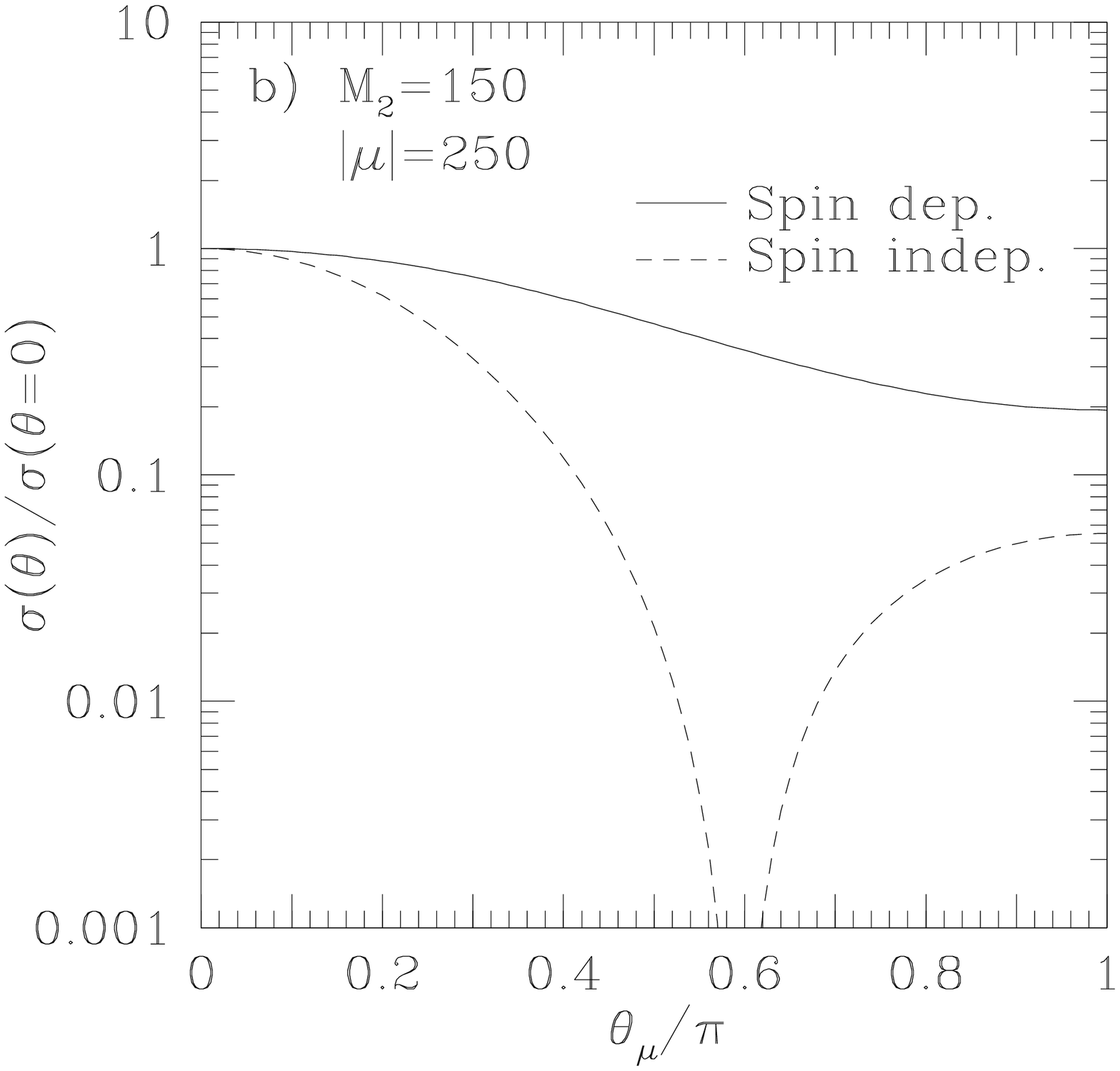,height=4.in} 
\caption{\label{fig:ge}For elastic scattering off of ${}^{73}$Ge, a) the absolute
value of the contributions to $f_n$ from individual quarks as a function of
$\thm$, and b) spin dependent (solid) and spin independent (dashed)
cross-sections as a function
of $\thm$, and normalized to the cross-sections at $\thm=0$.}
\end{center}
\end{figure}

The spin independent cross-section is dominant in much of the parameter
space for scattering off of heavy nuclei.  In Fig.\ref{fig:ge}, we consider the
scattering of neutralinos on $^{73}$Ge, for $M_2=150$ GeV and  $\mu=250$ GeV. 
In Fig.\ref{fig:ge}a, we show the
relative contributions to $f_n$.  The dominant contributions to $f_n$ come from
Higgs exchange, and from (\ref{a3}) and (\ref{fp}) one sees that contributions
from up-type quarks and down-type quarks are simply scaled by the appropriate
$f_T$'s.  The solid line shows the total $f_n$ (which is close to the total $f_p$), and
again one sees significant cancellations, near $\thm/\pi=0.6$.
The cancellation in the total $f_n$ occurs at a different place from that
of the individual contributions because of the relative signs of the
latter which are not shown. The signs of the up-type contributions
differ from those of the down-type (at $\thm \sim 0, \pi$) and both change
sign at $\thm/\pi \sim 0.6$. In Fig.~\ref{fig:ge}b,
we show by the dashed curve  the total value of the spin independent
cross-section
$\sigma_3(\thm)$, again normalized to
$\sigma_3(0)$. 
Here again, we see a strong dependence on the CP-violating phase
$\thm$, and for $\thm \sim 0.6$, there is again a strong cancellation in
the total scattering cross-section.  Note that in this case, the spin
dependent cross-section (shown by the solid curve) simply connects the
$\thm = 0$ and $\thm = \pi$ limits monotonically.

Finally, in Fig.~3, we show that the $\thm$ dependence of the
cross-sections does not always lead to cancellations and a diminishing of
the cross-section.  Indeed, while we generally do find a strong
dependence on $\thm$, in some cases this dependence leads to an
enhancement of the cross-section.  In Fig.~3, MSSM parameters were
chosen as $M_2 = 130 \gev$, $|\mu| = 110 \gev$, $\tan \beta = 2$ and $m_0
= 1500\gev$ (to satisfy the Higgs mass constraint).
As one can see, the dependence of the cross-section on
$\thm$ is not monotonic. The large variance in the cross-section from
$\thm = 0$ to
$\pi$ is largely due to the fact that the neutralino mass varies rapidly
for these parameters, from 26 to 70 GeV.  Note that for $\thm/\pi \la
0.5$,  the chargino  mass (which is also strongly dependent on $\thm$) is
below the current experimental constraint of about 91~GeV.

We have shown that the cross-sections for elastic neutralino-nucleus
scattering relevant for the detection of supersymmetric dark matter are
strongly dependent on the CP-violating phase $\thm$ associated with the
Higgs mixing mass $\mu$ in the MSSM.  For particular MSSM parameters, the
value of the the phase $\thm$ can lead to either strong cancellations or
in some cases enhancements to the cross-section and ultimately the
detection rate.  The full dependence on the MSSM parameters $M_2, \mu$,
and $\tan \beta$ as well as $A_0$ and its associated phase $\tha$ will be
presented elsewhere \cite{ffo2}.

\vskip 1in
\vbox{
\noindent{ {\bf Acknowledgments} } \\
\noindent  The work of A.F. and K.O. was supported in part by DOE grant
DE--FG02--94ER--40823.  The work of T.F. was supported in part by DOE   
grant DE--FG02--95ER--40896 and in part by the University of Wisconsin  
Research Committee with funds granted by the Wisconsin Alumni Research  
Foundation.}

\newpage

\begin{figure}
\begin{center}
\vspace*{-0.5in}
\hspace*{-0.0in}
\epsfig{file=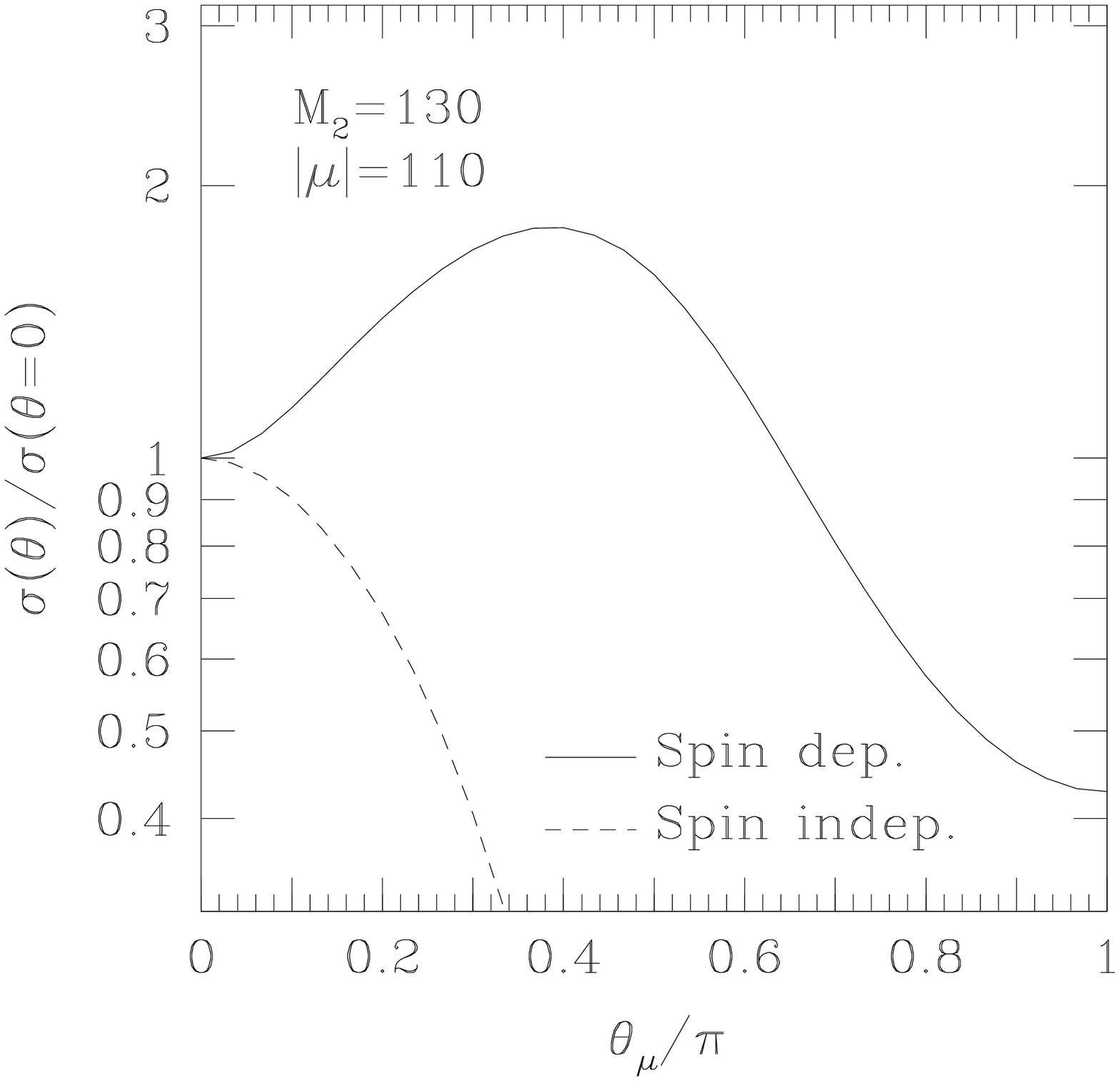,height=4.in} 
\caption{\label{fig:bu}An example of a $\thm$ dependence which shows an
enhancement in the  $\chi$-${}^{19}$F scattering  cross-section
rather than a cancellation. }
\end{center}
\end{figure}


\vskip .5in

\begin{thebibliography}{99}
\bibitem{gdyn} see e.g. J.R. Primack, in {\it Dark Matter in Astro- and
Particle Physics}, eds. H.V. Klapdor-Kleingrothaus and Y. Ramachers,
(World Scientific, Singapore, 1997) p. 97; and astro-ph/9707285. 
\bibitem{bbn} see e.g. K.A. Olive and D.N. Schramm, in the Reviews of
Particle Properties, Eur. Phys. J., {\bf C3} (1998) 1. 
\bibitem{efos2} J. Ellis, T. Falk, K.A. Olive and M. Schmitt,
Phys. Lett. {\bf B413} (1997)  355.
\bibitem{efgos} J. Ellis, T. Falk, G. Ganis, K.A. Olive and M. Schmitt,
Phys. Rev. {\bf D58} (1998) 095002.
\bibitem{someexperiment} see e.g. M. Maggi, to be published in the
Proceedings of the XXXIIIrd Rencontres de Moriond, March 1998.
\bibitem{jkg} G. Jungman, M. Kamionkowski, and K. Griest, Phys. Rep.
{\bf 267} (1996) 195.
\bibitem{sw} M. Srednicki and R. Watkins, Phys. Lett. {\bf B225} (1989)
140.

\bibitem{g} K. Griest, Phys. Rev. {\bf D38} (1988) 2357.


\bibitem{dgh}M. Dugan, B. Grinstein and L. Hall, Nucl. Phys. {\bf B255}, 413
(1985).
\bibitem{dn}J. Ellis, S. Ferrara, and D.V. Nanopoulos, Phys. Lett.
{\bf 114B} (1982) 231; W. Buchm\"{u}ller and D. Wyler, Phys. Lett.
{\bf 121B} (1983) 321;
J. Polchinski and M. Wise, Phys. Lett. {\bf 125B}  (1983) 393;
F. del Aguila, M. Gavela, J. Grifols, and A. Mendez, Phys. Lett. {\bf 126B}
(1983) 71; D.V. Nanopoulos and M. Srednicki, Phys. Lett. {\bf 128B}
(1983) 61. 

\bibitem{pil1}A. Pilaftsis,  Phys. Rev. {\bf D58} (1998) 096010, and 
Phys. Lett. {\bf B435} (1998) 88.

\bibitem{fkosi}T. Falk, K.A. Olive and M. Srednicki, Phys. Lett. {\bf 354}
(1995) 99.

\bibitem{fko1}T. Falk and K.A. Olive, Phys. Lett. {\bf 375} (1996) 196.


\bibitem{fko2}T. Falk and K.A. Olive, hep-ph/9806236.

\bibitem{gw}R. Garisto and J.D. Wells, Phys. Rev. {\bf D55}
(1997) 1611.


\bibitem{ffo2} T. Falk, A. Ferstl, and K.A. Olive, in preparation.

\bibitem{smc} D. Adams et al. (The Spin Muon Collaboration), Phys. Lett. {\bf B329} (1994)
 399.

\bibitem{svz} M. A. Shifman, A. I. Vainshtein, and V. I. Zakharov, Phys. Lett. {\bf B78}
(1978) 443; A. I. Vainshtein,  V. I. Zakharov, and M. A. Shifman Usp. Fiz. Nauk {\bf 130}
(1980) 537. 

\bibitem{gls}J. Gasser, H. Leutwyler, and M. E. Sainio, Phys. Lett. {\bf B253} (1991) 252.

\bibitem{dn1}M. Drees and M. M. Nojiri, Phys.Rev. {\bf D47} (1993) 4226. 

\bibitem{dn2}M. Drees and M. M. Nojiri, Phys.Rev. {\bf D48} (1993) 3483. 

\bibitem{ef} J. Ellis and R. Flores, Phys. Lett. {\bf B300}
(1993) 175.

\bibitem{higgs} see. e.g. L3 Collaboration, CERN Preprint CERN-EP98-72
(1998).

\end{thebibliography}
\end{document}